\documentclass[aps,prd,twocolumn,nofootinbib]{revtex4-2}

\usepackage[T1]{fontenc}
\usepackage{lmodern}
\usepackage{amsmath,amssymb,bm}
\usepackage{hyperref}
\hypersetup{colorlinks=true,citecolor=blue,linkcolor=blue,urlcolor=blue}

\newcommand{\Tr}{\operatorname{Tr}}
\newcommand{\diag}{\operatorname{diag}}
\newcommand{\one}{\mathbf 1}

\begin{document}

\title{Metric Signature as an Auxiliary Order Parameter: A Causal Viability Criterion}
\author{Miguel Bermudez}
\email{miguel.bermudez@u-paris.fr}
\affiliation{Universit\'e Paris Cit\'e - Sorbonne Universit\'e - CNRS,\\
IMJ-PRG, F-75013 Paris, France}
\date{\today}

\begin{abstract}
We describe a branchwise formulation of metric gravity in which the spacetime
signature is encoded by a real symmetric internal field $H_{ab}$ through
$g_{\mu\nu}=e_\mu{}^aH_{ab}e_\nu{}^b$.  An $O(4)$-invariant algebraic
potential has non-degenerate minima satisfying $H^2=\one$, with one vacuum
orbit for each inertia $(p,q)$.  Within a connected region where $H$ is
non-degenerate, Sylvester's law permits a change of variables that places all
derivatives in the metric sector.  The remaining components of $H$ are
auxiliary, while the flat directions along its $O(4)$ orbit are gauge.  We
derive this statement directly from the coupled tetrad and $H$ equations.
As a diagnostic example, we study $R+\alpha R^2$ gravity around the five flat
signature classes in four dimensions.  For fixed universal couplings, only one
Lorentzian branch has the standard graviton sign, a non-tachyonic scalaron, and
a hyperbolic initial-value problem.  Euclidean and split-signature branches do
not define ordinary causal phases.  Conditional on the existence of persistent
observers, this supplies an anthropic interpretation of the causal viability
criterion.  It is not a dynamical theory of transitions or a probability
measure on the branches: any real transition must encounter $\det H=0$, where
the branchwise description fails.
\end{abstract}

\maketitle

\section{Introduction}

The Lorentzian signature of spacetime is normally part of the kinematical
definition of general relativity.  Nevertheless, metrics of different inertia
are equally natural from the standpoint of differential geometry.  Euclidean
boundary conditions in quantum cosmology, real signature-changing geometries,
and complex spacetime metrics in gravitational path integrals have long been studied
\cite{HartleHawking,Ellis,DrayHellaby,WittenComplex}.  This raises a modest but
sharp question: can the different signatures be represented as vacuum sectors
of a single covariant field theory without adding a new propagating signature
mode?

We answer this question branchwise.  Introduce a tetrad $e_\mu{}^a$ and a real
symmetric internal field $H_{ab}$, and define
\begin{equation}
  g_{\mu\nu}=e_\mu{}^a H_{ab}e_\nu{}^b .
  \label{eq:composite-metric}
\end{equation}
The inertia of $H$ is then the signature of $g$ whenever $e$ and $H$ are
non-degenerate.  A potential for $H$ can place Euclidean, Lorentzian, and split
signatures among the stationary sectors of the same action.  Percacci identified
the soldering form and fibre metric as gravitational order parameters; Girelli,
Liberati, and Sindoni obtained Lorentzian propagation from a Riemannian scalar-field
background; and Bondarenko and Zubkov promoted an internal tangent-space metric in
Riemann-Cartan gravity \cite{Percacci,Girelli,Bondarenko}.

Unlike the closest of these constructions, Ref.~\cite{Bondarenko}, the present
model has neither an independent connection nor a derivative term for the
internal metric.  The curvature sector is an ordinary metric theory and $H$ is
strictly auxiliary away from degeneracy.  The new question is therefore not
whether a signature mode propagates, but whether algebraically degenerate
signature sectors support equally viable metric dynamics.

The central point of this note is not that the theory dynamically crosses
between these sectors.  Rather, it is that $H$ can consistently serve as an
\emph{auxiliary order parameter}: in every fixed non-degenerate branch, the
propagating phase space is that of the chosen metric theory.  We make the
field-equation argument explicit, identify the gauge and auxiliary directions
of $H$, and then use $R+\alpha R^2$ gravity as a simple spectral test.  This
separates three logically distinct statements: the existence of signature
vacua, the absence of additional local modes in a fixed branch, and the
viability of the metric excitations supported by that branch.  The
last step admits a conditional, anthropic interpretation.  An observer capable
of recording a history and inferring local laws presupposes a phase with a
stable causal structure and predictive evolution.  Conditioning on the
existence of such observers can therefore eliminate signature sectors even
when the algebraic potential assigns them equal vacuum value.

\section{Auxiliary signature field}

\subsection{Action and symmetry}

Let $L_g[g]$ be any local diffeomorphism-invariant metric Lagrangian.  We
consider
\begin{equation}
 S[e,H]=\int d^4x\,\sqrt{|g|}\left[L_g(g)-V(H)\right],
 \qquad g=eHe^T ,
 \label{eq:action}
\end{equation}
where the sign of $V$ is the conventional Lorentzian one.  None of the
branchwise degree-of-freedom conclusions below depends on this sign.  The
field $H$ is taken to be dimensionless, so that $\lambda$ below has mass
dimension four.  A minimal illustrative
choice is
\begin{equation}
 V(H)=\frac{\lambda}{4}\Tr\bigl(H^2-\one\bigr)^2,
 \qquad \lambda>0 .
 \label{eq:potential}
\end{equation}
This is not the unique even quartic $O(4)$ invariant; it is chosen because its
non-degenerate minima have a particularly transparent form.

The action is invariant under diffeomorphisms and under the local internal
transformations
\begin{equation}
 e\longmapsto e\Lambda^T,
 \qquad H\longmapsto\Lambda H\Lambda^T,
 \qquad \Lambda(x)\in O(4).
 \label{eq:o4}
\end{equation}
The non-degenerate minima of \eqref{eq:potential} satisfy
\begin{equation}
 H^2=\one .
 \label{eq:minima}
\end{equation}
Up to $O(4)$ conjugation, they are represented by
\begin{equation}
 H_{p,q}=\diag(\underbrace{-1,\ldots,-1}_{p},
                     \underbrace{+1,\ldots,+1}_{q}),
 \qquad p+q=4 .
 \label{eq:representatives}
\end{equation}
Thus there are five vacuum orbits, labelled by $(p,q)$.  Each orbit is the
coset $O(4)/[O(p)\times O(q)]$.  Its points are gauge-equivalent, rather than a
continuous family of physically distinct signatures.

The complete algebraic stationarity equation is
\begin{equation}
 H(H^2-\one)=0 .
 \label{eq:all-stationary}
\end{equation}
It also admits degenerate stationary points with zero eigenvalues.  Equation
\eqref{eq:minima} follows from \eqref{eq:all-stationary} in the non-degenerate
sector and selects the absolute minima of the algebraic function $V$.

\subsection{Branchwise separation of variables}

\paragraph{Branchwise auxiliary-field statement.}
For every non-degenerate signature branch and every metric Lagrangian
$L_g[g]$, the coupled theory \eqref{eq:action} has the same local propagating
degrees of freedom as $L_g$.  The components of $H$ transverse to its $O(4)$
orbit are auxiliary, and the tangent components are gauge.  The statement is
local in field space and does not extend through $\det H=0$.

Fix a connected region on which $H$ is non-degenerate and has inertia $(p,q)$.
Sylvester's law gives, locally,
\begin{equation}
 H=S H_{p,q}S^T,
 \qquad S\in GL(4,\mathbb R),
 \label{eq:sylvester}
\end{equation}
with $S$ defined up to right multiplication by $O(p,q)$.  Defining
\begin{equation}
 e'=eS
 \label{eq:new-tetrad}
\end{equation}
gives
\begin{equation}
 g=e'H_{p,q}e'^T .
 \label{eq:fixed-internal-metric}
\end{equation}
Consequently, every derivative in the curvature action can be expressed in
terms of $e'$ alone.  This is a change of variables, not an enlargement of the
gauge group from $O(4)$ to $GL(4)$.  The variables encoded by $S$ remain in
$V(SH_{p,q}S^T)$, but they enter without derivatives and are therefore
auxiliary.

The same conclusion follows directly from the equations of motion.  Define
the symmetric metric Euler derivative, including the contribution of the
measure multiplying $V$, by
\begin{equation}
 \mathcal E^{\mu\nu}:=
 \frac{2}{\sqrt{|g|}}
 \left.\frac{\delta S}{\delta g_{\mu\nu}}\right|_H .
 \label{eq:metric-euler}
\end{equation}
Using
$\delta g_{\mu\nu}=2\delta e_{(\mu}{}^aH_{ab}e_{\nu)}{}^b$
in the tetrad variation gives
\begin{equation}
 \mathcal E^{\mu\nu}H_{ab}e_\nu{}^b=0 .
 \label{eq:tetrad-eom}
\end{equation}
For invertible $e$ and $H$, it implies $\mathcal E^{\mu\nu}=0$.  The $H$
equation is
\begin{equation}
 \frac12e_\mu{}^{a}e_\nu{}^{b}\mathcal E^{\mu\nu}
 -\frac{\partial V}{\partial H_{ab}}=0 .
 \label{eq:h-eom}
\end{equation}
Combining \eqref{eq:tetrad-eom} and \eqref{eq:h-eom} yields the algebraic
condition
\begin{equation}
 \frac{\partial V}{\partial H}
 =\lambda H(H^2-\one)=0 .
 \label{eq:potential-eom}
\end{equation}
The $H$ equation is therefore not
algebraic before the metric equation is used, but the coupled system contains
no additional propagating $H$ mode.

This structure can also be seen from the Hessian of the potential at
$H_{p,q}$.  To quadratic order,
\begin{equation}
 \delta^2V\ \propto\
 \Tr\!\left(\{H_{p,q},\delta H\}^2\right).
 \label{eq:hessian}
\end{equation}
The $pq$ components mixing the positive and negative eigenspaces are null
directions.  They coincide with the gauge directions generated outside the
stabilizer $O(p)\times O(q)$, since
\begin{equation}
 \dim O(4)-\dim O(p)-\dim O(q)=pq .
\end{equation}
The remaining
\begin{equation}
 10-pq=\frac{p(p+1)}2+\frac{q(q+1)}2
 \label{eq:auxiliary-count}
\end{equation}
components are transverse to the gauge orbit and are fixed algebraically.
Hence the physical
local degrees of freedom are precisely those of $L_g[g]$, subject to the usual
constraints of that metric theory.

The qualification ``branchwise'' is essential.  The inertia of a real
symmetric matrix cannot change without an eigenvalue passing through zero.
At $\det H=0$, the metric is degenerate and the standard curvature action is
generally undefined.  The present theory therefore classifies fixed-signature
phases; by itself it is not a regular classical theory of signature change.

\section{Causal and spectral viability of the signature branches}

To illustrate how the metric sector distinguishes the vacuum orbits, consider
a quadratic $f(R)$ truncation without the massive spin-2 ghost on the target
Lorentzian branch,
\begin{equation}
 L_g=\sigma R+\alpha R^2,
 \qquad \sigma>0 .
 \label{eq:starobinsky}
\end{equation}
We deliberately omit a Weyl-squared term: a nonzero coefficient would add the
usual massive spin-2 pole with the opposite residue
\cite{Stelle,Hindawi,Salvio}.  The theory \eqref{eq:starobinsky} propagates the
two graviton helicities and one scalaron.  Around the $(-+++)$ Minkowski
branch, the scalaron mass is
\begin{equation}
 m_0^2=\frac{\sigma}{6\alpha},
 \label{eq:scalaron-mass}
\end{equation}
up to the normalization convention for the quadratic term.  Thus
$\alpha>0$ gives the standard graviton residue and a non-tachyonic scalaron
\cite{Starobinsky,Hindawi,Salvio}.
Equivalently, introducing
\begin{equation}
 \Phi=f'(R)=\sigma+2\alpha R
\end{equation}
rewrites the curvature sector as
\begin{equation}
 L_g=\Phi R-\frac{(\Phi-\sigma)^2}{4\alpha} .
 \label{eq:scalar-tensor-form}
\end{equation}
The conditions $f'(0)>0$ and $m_0^2=f'(0)/[3f''(0)]>0$ reproduce
$\sigma>0$ and \eqref{eq:scalaron-mass}.

The two Lorentzian vacuum orbits are related by $g\mapsto-g$.  In four
dimensions,
\begin{equation}
 R[-g]=-R[g],\qquad R^2[-g]=R^2[g].
 \label{eq:sign-reversal}
\end{equation}
For fixed $(\sigma,\alpha)$, the opposite Lorentzian branch therefore has the
effective replacement $\sigma\mapsto-\sigma$ but not
$\alpha\mapsto-\alpha$.  It is not merely a change of sign convention within
this enlarged configuration space: the action itself distinguishes the two
branches.  With $\sigma>0$ and $\alpha>0$, the branch $(1,3)$ has the standard
massless residue and $m_0^2>0$, whereas the sign-reversed branch fails this
combined test.  In the $(-+++)$ convention the sign-reversed branch is
described by
\begin{equation}
 \widetilde f(R)=-\sigma R+\alpha R^2,
\end{equation}
so that
\begin{equation}
 \widetilde f'(0)=-\sigma<0,
 \qquad
 \widetilde m_0^2=-\frac{\sigma}{6\alpha}<0 .
 \label{eq:opposite-branch-test}
\end{equation}
Multiplying the complete curvature action by an overall minus sign can reverse
all residues, but it does not change the ratio
$\widetilde f'(0)/\widetilde f''(0)$ and therefore does not remove the
tachyonic scalaron.  The distinction between the two branches is thus already
present in the pure metric sector whenever both the odd term $R$ and the even
term $R^2$ are retained.  Matter is required for a phenomenologically complete
theory, but not for this branchwise diagnostic.

For the Euclidean classes $(0,4)$ and $(4,0)$, the linearized equations are
elliptic.  Such saddles may be relevant to a Euclidean functional integral,
but they do not define a real-time Cauchy evolution and should not be called
dynamically unstable on that ground alone.  The split class $(2,2)$ is
ultrahyperbolic.  Generic local data on a codimension-one hypersurface do not
define the usual unconstrained Cauchy problem; well-posed evolution can be
recovered only for restricted data satisfying a nonlocal constraint
\cite{John,CraigWeinstein}.  The branchwise classification is therefore:
\begin{itemize}
 \item $(1,3)$: hyperbolic, with $f'(0)>0$ and $m_0^2>0$;
 \item $(3,1)$: hyperbolic, but with $f'(0)<0$ and $m_0^2<0$;
 \item $(0,4)$ and $(4,0)$: elliptic, with no real-time evolution;
 \item $(2,2)$: ultrahyperbolic, without the usual unconstrained local Cauchy
 problem.
\end{itemize}

The condensed-matter perspective supplies a useful proof of principle for
interpreting signature as a property of an effective excitation sector.
Volovik emphasizes that metric and tetrad structures can emerge from
inequivalent microscopic vacua \cite{VolovikReview}.  More specifically,
Nissinen and Volovik show that the quadratic action of the magnon-BEC
pseudo-Goldstone mode in polar $^3$He changes from an effective Minkowski
metric in the stable convex regime to a Euclidean metric when the interaction
becomes concave; the latter regime has complex mode frequencies and is
dynamically unstable \cite{NissinenVolovik}.  This is neither a microscopic
model of spacetime nor an auxiliary-field construction of its metric.  It
nevertheless provides a concrete analogue of the principle used here: the
signature inferred by low-energy excitations and the stability of their
propagation are linked.

The conclusion is a selection by \emph{causal and spectral viability}: among the
algebraically degenerate signature vacua, one Lorentzian orbit supports the
standard low-energy propagation of the metric modes for a fixed choice of
couplings.  This does not yet provide a probability distribution over the
vacua, a relaxation mechanism, or a transition amplitude.  Those stronger
questions require a microscopic dynamics near $\det H=0$ or a separately
defined complex integration cycle.

\subsection{Anthropic conditioning}

Let $B_{p,q}$ denote a signature branch and let $\mathcal O$ denote the
existence of persistent observers, understood minimally as physical systems
capable of storing records and undergoing predictable local evolution.  The
proposal is the conditional statement
\begin{equation}
 P(B_{p,q}\mid\mathcal O)\propto
 P(\mathcal O\mid B_{p,q})P(B_{p,q}).
 \label{eq:anthropic-conditioning}
\end{equation}
The present model does not determine the prior weights $P(B_{p,q})$.  It only
supplies a necessary-condition filter.  For an elliptic Euclidean
sector, an ordinary real-time history is absent; for a generic ultrahyperbolic
sector, evolution from codimension-one initial data is not predictive; and for
a Lorentzian sector with ghosts or tachyons, persistent structures are not
expected to be stable.  In this minimal sense,
\begin{equation}
 P(\mathcal O\mid B_{p,q})\simeq0
 \label{eq:zero-observer-weight}
\end{equation}
for branches failing the causal and spectral tests, in the idealized effective
description used here.  If a single Lorentzian
branch survives, conditioning on $\mathcal O$ selects it independently of the
unknown nonzero prior assigned to that branch.

Equation \eqref{eq:zero-observer-weight} should be read as an idealized
necessary-condition statement, not as a calculation of observer abundance.
The construction neither defines observers microscopically nor claims to
derive their absolute probability.  Its content is that stable causal
propagation is a prerequisite for any observer described within the effective
field theory.

\section{Discussion}

The construction provides a compact way to place all metric signatures in a
single field space while preserving the local propagating content of an
arbitrary metric theory inside each non-degenerate branch.  Its essential
ingredients are algebraic: an internal symmetric field records the inertia,
the potential selects representative vacuum orbits, and the coupled field
equations make $H$ auxiliary.  The example $R+\alpha R^2$ then shows that
degeneracy of the algebraic potential need not imply equivalence of the
propagating phases.  In the illustrative $R+\alpha R^2$ sector, the coexistence
of a term odd and a term even under $g\mapsto-g$ is enough to distinguish the
two Lorentzian branches spectrally.

This leads to a limited anthropic conclusion.  The potential does not prefer
one inertia class, but observers can occur only in a branch supporting stable,
predictive causal evolution.  The observed Lorentzian signature may therefore
be understood as the result of conditioning on the existence of observers in
the space of signature sectors.  The argument is insensitive to the unknown
prior weights only when all competing branches have vanishing observer weight;
if more than one viable Lorentzian branch survives after the full field content
is included, an actual measure would again be required.

Two limitations are structural rather than technical.  First, the model does
not make $H$ relax: a strictly auxiliary field obeys a constraint rather than
a wave or dissipative equation.  Second, real signature change reaches a
degenerate metric where the starting action ceases to apply.  Accordingly, the
present result should be read as a theory of signature \emph{sectors}, their
spectra, and their conditional viability.  A theory of dynamical transitions
or of prior probabilities would require additional
input, for example a UV completion resolving degenerate metrics or a precise
holomorphic action and integration cycle for complex metrics.

\end{document}